\documentclass[conference]{IEEEtran}
\IEEEoverridecommandlockouts
\usepackage{cite}
\usepackage{amsmath,amssymb,amsfonts}
\usepackage{algorithmic}
\usepackage{graphicx}
\usepackage{textcomp}
\usepackage{xcolor}

\usepackage{epsfig}
\usepackage{amsfonts}
\usepackage{listings}
\usepackage{lmodern}
\usepackage{mathtools}
\usepackage{graphicx}
\usepackage{textcomp}
\usepackage{epstopdf}
\usepackage{enumerate}
\usepackage{color}
\usepackage{cleveref}
\usepackage{flushend}

\newtheorem{theo}{Theorem}
\newtheorem{lem}{Lemma}
\newtheorem{cor}{Corollary}

\newtheorem{propo}{Proposition}
\newtheorem{defi}{Definition}

\newtheorem{assum}{Assumption}

\DeclareMathOperator{\xun}{x_{\textup{un}}}
\DeclareMathOperator{\bx}{\bold{x}}

\DeclareMathOperator{\vdes}{v_{\textup{des}}}
\DeclareMathOperator{\vsafe}{v_{\textup{safe}}}

\def\BibTeX{{\rm B\kern-.05em{\sc i\kern-.025em b}\kern-.08em
    T\kern-.1667em\lower.7ex\hbox{E}\kern-.125emX}}
\begin{document}

\title{Safe Sliding Mode Control in Position for Double Integrator Systems\\

}

\author{\IEEEauthorblockN{1\textsuperscript{st} Marco A. Gomez}
\IEEEauthorblockA{\textit{Department of Automatic Control} \\
\textit{Cinvestav}\\
Mexico City, Mexico \\
marco.gomez@cinvestav.mx}
\and
\IEEEauthorblockN{2\textsuperscript{nd} Christopher D. Cruz-Ancona}
\IEEEauthorblockA{
\textit{Tecnologico de Monterrey, }\\
\textit{School of Engineering and Sciences,}\\
Nuevo Leon, Mexico \\
christopher.cruz.ancona@tec.mx}
}

\maketitle

\begin{abstract}
	We address the problem of robust safety control design for double integrator systems. We show that, when the constraints are defined only on position states, it is possible to construct a safe sliding domain from the dynamic of a simple integrator that is already safe. On this domain, the closed-loop trajectories remain robust and safe against uncertainties and disturbances. Furthermore, we design a controller gain that guarantees convergence to the safe sliding domain while avoiding the given unsafe set.  The concept is initially developed for first-order sliding mode and is subsequently generalized to an adaptive framework, ensuring that trajectories remain confined to a predefined vicinity of the {sliding domain, outside} the unsafe region.
\end{abstract}

\begin{IEEEkeywords}
Sliding Mode Control, Safety control, double integrator systems
\end{IEEEkeywords}

	\section{INTRODUCTION}

The design of control algorithms aiming at robustness and safety properties of different classes of systems has been a central problem in the last decades. Although several approaches to construct safety controllers can be adopted, design of \textit{robust safety} controllers \cite{KolathayaAmes2018} remains a main issue that has not been fully addressed. An instance of this is the existing gap between Sliding Mode Control (SMC) and safety-critical control based on the notion of Control Barrier Functions (CBF), which has proved effectiveness in several robotic applications \cite{Amesetal2016,Molnar2021model}. A first step to filling this gap was presented in \cite{gomez2022cce,Gomezancona2025}, where the notion of Safe Sliding Manifold (SSM) was introduced. Roughly speaking, a SSM is a manifold in which the trajectories of the closed-loop system slide without entering any unsafe set while {maintaining robustness against bounded uncertainties and disturbances.}

The approach adopted in \cite{Gomezancona2025} to construct a SSM relies on the gradient of a Lyapunov-like function, used for safety controller design in \cite{RomdlonyJayawardhana2016}. The proposal faces two main problems: (i) it assumes the existence (and requires the construction) of both a Control Lyapunov Function (CLF) and  a CBF; and (ii) it demands a planned trajectory design in the state space to ensure that the closed-loop trajectories reach the SSM without entering the unsafe set; this means that, without a priori planned trajectory, it is not possible to guarantee a safe reaching phase. 

In this note, we consider the problem of robust safety controller design for systems described by a double integrator, with uncertainties and disturbances, of the form
\begin{equation}
	\label{ec:sys_di}
	\ddot x(t)=G(\bx)\left((I+\Delta_b(t,\bx))u(t)+\delta(t,\bx)\right),\:t\geq 0,
\end{equation}
where  $G:\mathbb{R}^{2n}\rightarrow \mathbb{R}^{n\times n}$ is  locally Lipschitz and is non-singular, $u(t)\in \mathbb{R}^n$ is the input function,  
$x\in \mathcal{X}_1\subseteq \mathbb{R}^n $ and $\dot x\in \mathcal{X}_2\subseteq \mathbb{R}^n$. The corresponding state space is denoted by $\mathcal{X}:=\mathcal{X}_1 \times \mathcal{X}_2$, and to be more concise in the notation, we define any vector on $\mathcal{X}$ as $\bx=\begin{pmatrix}
	x^T & \dot x^T
\end{pmatrix}^T$ and the corresponding vector of initial conditions as $\bold{x_0}:=~\begin{pmatrix}
	x_0^T & \dot x_0^T
\end{pmatrix}^T$. The functions $\Delta_b:\mathbb{R}_{\geq 0}\times \mathbb{R}^{2n}\rightarrow \mathbb{R}^{n\times n}$ and $\delta:\mathbb{R}_{\geq 0}\times \mathbb{R}^{2n}\rightarrow \mathbb{R}^{n} $, measurable in $t$ and continuous in $\bx$, characterize uncertain and disturbance terms.

We depart from the definition of safe sliding domain introduced in \cite{Gomezancona2025} to construct a robust controller that preserves safety of the trajectories of \eqref{ec:sys_di} in position, during both the sliding and reaching phase.  We do not require a CLF, but only a safety controller for a simple integrator, obtained within the framework developed in \cite{Amesetal2016}. 

To be more precise, we now present the definitions that underpin the main results. Hereafter,  the solution of the corresponding state space equation of \eqref{ec:sys_di}, which emanates from a specific initial condition $\bx_0$, is denoted by $\bx(t,\bx_0)$. Since the controller $u$ is considered to be a discontinuous function of the state, the solutions are understood in the sense of Filippov \cite{Filippov2013}, i.e. $\bx(t)$ is an absolutely continuous function satisfying the corresponding differential inclusion almost everywhere. 

	\begin{defi}
	\label{def:maindefi}
	Let $\mathcal{M}$ be a differentiable manifold of dimension $r<2n$ in $\mathcal{X}$ and $\mathcal{C}$ be a subset of $\mathcal{X}$. 
	\begin{enumerate}
		\item A non-empty subset $\mathcal{R}\subseteq \mathcal{M}$, relatively open in $\mathcal{M}$, is a sliding domain for \eqref{ec:sys_di} if for any $y\in \mathcal{R}$ there exists $\epsilon>0$ such that, if $\|\bx_0-y\|<\epsilon$, $\bx_0\in \mathbb{R}^{2n}$, then it can be found $T(\bx_0)$ such that $\bx(T(\bx_0))\in \mathcal{R}$ and  $\lim_{x_0\rightarrow y}T(\bx_0)=0$.
		\item The set $\mathcal{R}$ is a safe sliding domain with respect to  $\mathcal{C}$ for \eqref{ec:sys_di}  if $\mathcal{R}$  is a sliding domain and  $\bx(t,\bx_0)\in \mathcal{C}$ for all $t\geq T(\bx_0)$.
	\end{enumerate}
\end{defi}

\begin{figure}[htbp!]
	\centering
	\includegraphics[width = 0.25 \textwidth]{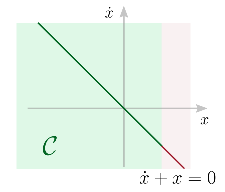}
	\caption{A simple construction of a safe sliding domain with respect to a set $\mathcal{C}$.}
	\label{fig:domain}
\end{figure}

The set $\mathcal{R}$ is a sliding manifold for \eqref{ec:sys_di} if $\mathcal{R}$ is a sliding domain and $\mathcal{R}=\mathcal{M}$. The first part of the above definition is presented in \cite{Utkin2020road}, while the second in \cite{Gomezancona2025}. Throughout this note, we consider $$\mathcal{C}=\mathcal{C}_1\times \mathcal{X}_2=\left\{(x,\dot x): x\in \mathcal{C}_1, \dot x\in \mathcal{X}_2\right\}.$$ For this case, to be a safe sliding domain, $\mathcal{R}$ does not require to satisfy constraints on $\mathcal{X}_2$.  Consider, for example, system \eqref{ec:sys_di} with $n=1$, and $\mathcal{C}_1=\left\{x\in\mathbb{R}: x\in [a, b] \right\}$ for any real numbers $a$ and $b$.   A manifold $\mathcal{S}$ can be constructed as $\left\{\bx \in \mathcal{X}: \dot x+ x=0\right\}$, which has a subset contained within $\mathcal{C}$; see Figure \ref{fig:domain}. If the trajectories of \eqref{ec:sys_di} (in closed-loop with a given controller) reach any point of $\mathcal{S}\cap \mathcal{C}$ in a finite time $T(\bx_0)$ and remain therein after that, then $\mathcal{S}$ is a safe sliding domain with respect to the set $\mathcal{C}$. 

Whereas the selection of $\mathcal{S}$ in this case is conventional, it is not in higher-dimensional state space cases. In addition, notice that with a standard sliding mode controller the initial position can satisfy $x_0\in \mathcal{C}_1$, but $x(t,\bx_0)\notin \mathcal{C}_1$ for some $t>0$ during the reaching phase. 

In line with Definition \ref{def:maindefi} we have:
\begin{defi}
	System \eqref{ec:sys_di} is safe with respect to $\mathcal{C}$ if $\bx_0\in \mathcal{C}\implies \bx(t,\bold{x_0})\in \mathcal{C}\:\forall t\geq 0$.
\end{defi}

The rest of the paper is organized as follows. In Section \ref{sec:ssm} we present the main results, namely, we provide a first-order sliding mode safety control and then we generalize it to an adaptive framework, ensuring that trajectories remain confined to a predefined vicinity of the sliding domain, outside the unsafe region. In Section \ref{sec:example} we illustrate the theoretical findings with a numerical example and, in Section \ref{sec:conclusion}, we conclude with some final remarks. We adopt standard notation and clarify it whenever required.

We assume the following on the uncertainties and disturbances:

\begin{assum}
	\label{ass:delta}
	There exists a function $d:\mathbb{R}_{\geq 0}\times \mathbb{R}^{2n}\rightarrow \mathbb{R}_{\geq 0}$ such that $\|\delta(t,\bx)\|\leq d(t,\bx)$ for all $(t,\bx)\in \mathbb{R}_{\geq 0}\times \mathbb{R}^{2n}$.
\end{assum}

\begin{assum}
	\label{ass:inputmatrix}
	For some positive real number $\gamma$, $\|\Delta_b(t,\bx)\|\leq \gamma <1$ for all $(t,\bx)\in \mathbb{R}_{\geq 0}\times \mathbb{R}^{2n}$. Moreover, there exists a known $\mu>-1$ such that
	\begin{multline*}
		\lambda_{\min}\left(\frac{1}{2}(G(\bx)\Delta_b(t,\bx)G^{-1}(\bx)\right.\\\left.+(G(\bx)\Delta_b(t,\bx)G^{-1}(\bx))^T)\right)\geq \mu \:\:\forall (t,\bx)\in \mathbb{R}_{\geq 0}\times \mathbb{R}^{2n}.
	\end{multline*}
\end{assum}

	\section{Robust safety control design in position}
\label{sec:ssm}

The results are build on the proposed manifold 
\begin{equation*}
	\mathcal{S}:=\left\{\bx\in \mathcal{X}:\sigma(\bx)=0\right\},
\end{equation*}
where 
\begin{equation}
	\label{ec:sliding_var}
	\sigma(\bx)= \dot x-v(x),
\end{equation}
with $v:\mathcal{X}_1\rightarrow \mathbb{R}^n$  a safety-critical control of the \emph{simple integrator} $\dot x = v(x)$. As we shall (formally) prove, such $v$ induces  $\mathcal{S}$ to be a safe sliding domain  with respect to $\mathcal{C}$. The manifold $\mathcal{S}$ is  reminiscent of the early work by \cite{Guldner1995sliding}, where the gradient of potential fields for obstacle avoidance was considered instead of $v$.  It is worth mentioning that the advantages of using CBFs instead of potential fields for obstacle avoidance, as discussed in \cite{Singletary2021comparative}, are inherited by the presented approach.

For the sake of completeness, we provide here $v$ \cite{Amesetal2016}. 
\begin{defi} \cite{Amesetal2016,Singletary2021comparative}
	Let  $h:\mathcal{X}_1\rightarrow \mathbb{R}$ be  twice continuously differentiable function and  $\mathcal{C}_1$ be the  safe set defined as
	\begin{equation*}
		\mathcal{C}_1=\left\{x\in \mathcal{X}_1:h(x)\geq 0\right\},
	\end{equation*}
	which is closed.  Then, $h$ is a  Control Barrier Function (CBF) of $\dot x(t)=v(x)$   if $\nabla h(x)\neq 0$ for all $x\in \partial \mathcal{C}_1$ and there is an extended class $\mathcal{K}$ function \footnote{An extended class $\mathcal{K}$ function is a continuous function $\alpha:(-b,a)\rightarrow (-\infty,\infty)$ for some $a,b>0$ that is strictly increasing and $\alpha(0)=0$, see \cite{Amesetal2016}.} $\alpha$ such that for all $x\in \mathcal{C}_1$ there exists $v$ satisfying
	\begin{equation*}
		\dot h(x)= \nabla h(x) v(x)\geq -\alpha(h(x)).
	\end{equation*}
\end{defi}

In the standard definition, $h$ is not required to be twice continuously differentiable. Here, we assume that to guarantee smoothness of $\mathcal{S}$.
Given a CBF, a safety control can be constructed by 
\begin{equation*}
	\begin{split}
		v^*(x)=\underset{{v\in \mathbb{R}^n}}{\textup{argmin}}&\|v-\vdes(x)\|^2\\
		\text{s.t.}\:& \nabla h(x) v \geq -\alpha(h(x)),
	\end{split}
\end{equation*}
where $\vdes$ is a desired velocity. The QP  has the explicit solution \cite{Xu2015robustness}
\begin{equation}
	\label{ec:usafe_gen}
	v^{*}(x)=\vdes(x)+\vsafe(x),
\end{equation}
where
\begin{multline*}
	\vsafe(x)=\\\left\{\begin{array}{cc}
		-\frac{\nabla h(x)^T}{\nabla h(x)\nabla h(x)^T}\left(\nabla h(x)\vdes(x)+\alpha(h(x))\right), & x\in \Psi\\
		0, &\text{otherwise}
	\end{array}\right.,
\end{multline*}
with $\Psi:=\{x\in \mathbb{R}^n: \nabla h(x)\vdes(x)<-\alpha(h(x))\} $. Hereafter, for simplicity, we set $\alpha(h(x)))=\alpha h(x)$ with $\alpha>0$. As shown in \cite{Xu2015robustness}, the controller $v$ is locally Lipschitz continuous whenever $\vdes$ is and $\nabla h(x)\neq 0$ for an open set $\mathcal{D}$, containing to $\mathcal{C}_1$.   The result follows from composition properties of  Lipschitz  continuous functions and the fact that $\vsafe(x)=\nu_0(\nu_1(x))\nu_2(x)$, where 
\begin{equation*}
	\begin{split}
		\nu _0(z)&=\left\{\begin{array}{cc}
			0,& z\geq 0\\
			z,&z< 0
		\end{array}\right.,\\
		\nu_1(x)&=\nabla h(x)\vdes(x)+\alpha(h(x)),\\
		\nu_2(x)&=-\frac{\nabla h(x)^T}{\|\nabla h(x)\|^2}.
	\end{split}
\end{equation*}
Smoothness of $\mathcal{S}$ is guaranteed by $\nabla h(x)\neq 0$ on $\mathcal{D}$, and approximation of $\nu_0$ by the smooth function
\begin{equation*}
	\nu_s(z)=\left\{\begin{array}{cc}
		0,& z\geq 0\\
		\frac{z}{2}\left(1-\cos\left(\frac{z\pi}{s}\right)\right) & -s<z<0\\
		z,&z< -s
	\end{array}\right.,
\end{equation*}
which, indeed, allows to approximate   $\vsafe$ by
\begin{equation}
	\vsafe^s=\nu_s(\nu_1(x))\nu_2(x).
\end{equation}
Since  $\nu_s\rightarrow \nu_0$ as $s\rightarrow 0$, we have that $\vsafe^s\rightarrow \vsafe$ as $s\rightarrow 0$.  For the construction of \eqref{ec:sliding_var}, we consider then
\begin{equation*}
	v(x)=\vdes(x)+\vsafe^s(x),
\end{equation*}
with $s$ a sufficiently small parameter such that $v(x)$ assures safety of $\dot x(t)=v(x)$ with respect to $\mathcal{C}_1$. Notice that manifold smoothing is also addressed in  \cite{Guldner1995sliding}, albeit from a different perspective.

The section is split into two parts. A sliding mode control algorithm that ensures safety of the system in spite of disturbances is presented  in Subsection \ref{sec:smc}; and a modified  control gain within an adaptive framework that enables the relaxation of Assumption \ref{ass:delta}  is introduced in Subsection \ref{sec:barrierfunction}.

	\subsection{Safe sliding mode control}
\label{sec:smc}
Let the control
\begin{equation}
	\label{ec:control_smc}
	u(t,\bx)=-k(t,\bx)G^{-1}(\bold{x})\frac{\sigma(\bx)}{\|\sigma(\bx)\|},
\end{equation}
where 
\begin{equation}
	\label{ec:gaink}
	k(t,\bx)= \frac{\kappa+\rho(t,\bx)}{1+\mu},
\end{equation}
with $\kappa>0$ and 
\begin{equation*}
	\rho (t,\bx)=\|G(\bx)\|d(t,\bx)+\|\dot v(x)\|.
\end{equation*}

\begin{theo}
	\label{theo:safe_manifold}
	Let  Assumption \ref{ass:delta} and Assumption \ref{ass:inputmatrix} be satisfied. The set $\mathcal{S}$ is a safe sliding domain for \eqref{ec:sys_di} in closed loop with  \eqref{ec:control_smc}, with $T(\bx_0)=\frac{\sqrt 2\|\sigma(\bx_0)\|}{\kappa}$. 
\end{theo}
\textit{Proof.}
	We first prove that $\mathcal{S}$ is a sliding domain, and then that it is also safe. Let the Lyapunov function
	\begin{equation*}
		V(\sigma)=\frac{1}{2}\sigma^T(\bx)\sigma(\bx).
	\end{equation*}
	Since 
	\begin{equation*}
		\begin{split}
			\dot \sigma (\bx)=&\ddot x -\dot v(x)\\
			=&  G(\bx)\left((I+\Delta_b(t,\bx))u+\delta (t,\bx)\right)-\dot v(x) \\
			=&-k(t,\bx)\frac{\sigma(\bx)}{\|\sigma(\bx)\|}\\
			&-k(t,\bx)G(\bx)\Delta_b(t,\bx)G^{-1}(\bx)\frac{\sigma(\bx)}{\|\sigma(\bx)\|}+\delta_0(t,\bx),
		\end{split}
	\end{equation*}
	where $\delta_0(t,\bx)=G(\bx)\delta(t,\bx)-\dot v(x)$, the time-derivative of $V$ satisfies
	\begin{equation*}
		\begin{split}
			\dot V(\sigma)=&\sigma^T(\bx)\dot\sigma(\bx)\\
			=& -k(t,\bx)\|\sigma(\bx)\|\\
			&-k(t,\bx)\sigma^T(\bx)G(\bx)\Delta_b(t,\bx)G^{-1}(\bx)\frac{\sigma(\bx)}{\|\sigma(\bx)\|}\\
			&+\sigma^T(\bx)\delta_0(t,\bx).
		\end{split}
	\end{equation*}
	As $\sigma^T(\bx)R\sigma(\bx)=\sigma^T(\bx)R^T\sigma(\bx)$ for any matrix $R$, 
	\begin{equation*}
    \begin{split}
        		\sigma^T(\bx)G(\bx)&\Delta_b(t,\bx)G^{-1}(\bx)\sigma(\bx)\\
		=&\frac{1}{2}\sigma^T(\bx)(G(\bx)\Delta_b(t,\bx)G^{-1}(\bx)\\+&(G(\bx)\Delta_b(t,\bx)G^{-1}(\bx))^T)\sigma(\bx)
            \end{split}
	\end{equation*}
	and from the Rayleigh-Ritz inequality it follows that
	\begin{equation}
		\label{ec:ineq_V}
		\dot V(\sigma)\leq-\|\sigma(\bx)\|\left(k(t,\bx)(1+\mu)-\rho(t,\bx)\right)= -\kappa \|\sigma(\bx)\| .
	\end{equation}
	The solution of the above differential inequality yields
	\begin{equation*}
		2\left(V^{\frac{1}{2}}(\sigma(\bx(t))-V^{\frac{1}{2}}(\sigma(\bx_0))\right)\leq -\kappa t
	\end{equation*}
	from which it follows that
	\begin{equation}
		\label{ec:sigma_bound}
		0\leq \|\sigma(\bx(t))\|\leq \|\sigma(\bx_0)\|-\frac{\kappa}{\sqrt{2}}t.
	\end{equation}
	By continuity of $\sigma$, for any $\delta_s>0$ there exists $\varepsilon_s>0$ such that for any $y\in \mathcal{S}$ 
	\begin{equation}
		\label{ec:continuity_sigma}
		\|\bx_0-y\|<\varepsilon_s \Rightarrow \|\sigma(\bx_0)-\sigma(y)\|=\|\sigma(\bx_0)\|<\delta_s.
	\end{equation}
	From \eqref{ec:continuity_sigma}, $T(\bx_0)=\frac{\sqrt 2\|\sigma(\bx_0)\|}{\kappa}<\frac{\sqrt 2\delta_s }{\kappa}$, and  from \eqref{ec:sigma_bound},  $$0\leq \|\sigma(\bx(t))\|\leq  \delta_s -\delta_s=0$$ for $t\geq T(\bx_0)$, therefore $\|\sigma(\bx(t))\|=0$ for $t\geq T(\bx_0)$.  By \eqref{ec:continuity_sigma} again, for any $\delta_s'>0$ there is $\varepsilon_s>0$ such that 
	\begin{equation*}
		\|\bx_0-y\|<\varepsilon_s \Rightarrow |T(\bx_0)|<\delta_s',
	\end{equation*}
	where $\delta_s^{'}=\frac{\sqrt{2}}{\kappa}\delta_s$, and $\delta_s$ is arbitrary, meaning that  $T(\bx_0)\rightarrow 0$ as $\bx_0\rightarrow ~y$. Finally,  on $\bx\in \mathcal{S}$, the system dynamics satisfy $\dot x=v(x)$, which by construction of $v$ implies that $x(t,\bx_0)\in \mathcal{C}_1$, equivalently $\bx(t,\bx_0)\in \mathcal{C}$, for all $t\geq T(\bx_0)$. \IEEEQEDclosed

While trajectories are safe and robust on $\mathcal{S}$, nothing can be said on them during the reaching phase. In the following proposition we show that the constant $\kappa$ in control \eqref{ec:control_smc} can be tuned to  guarantee  robust safety convergence to the safe sliding domain.

\begin{propo}
	\label{propo:reaching}
	Let  Assumption \ref{ass:delta} and Assumption \ref{ass:inputmatrix} be satisfied and assume that the gradient of $h$ admits a maximum on $\mathcal{X}_1$, defined by $\eta:=\max_{x\in \mathcal{X}_1 }\|\nabla h(x)\|$. If 
	\begin{equation}
		\label{ec:kappa}
		\kappa= \frac{\alpha}{2}\|\sigma (\bx_0)\|+\alpha_c \eta,
	\end{equation}
	where $\alpha$ is a constant  taken from  the construction of \eqref{ec:usafe_gen} and $$\alpha_c=\frac{\|\sigma(\bx_0)\|^2+\beta}{2 h(x_0)},$$ with any $\beta>0$,  then the trajectories of  \eqref{ec:sys_di} in closed loop with  \eqref{ec:control_smc} satisfy $x(t,\bx_0)\in \mathcal{C}_1$ for $t\in[0,T(\bx_0))$ for any $\bx_0\in \mathcal{C}$.
\end{propo}

\textit{Proof.} 
	We draw from \cite{Molnar2021model} and  consider the function
	\begin{equation*}
		h_c(\bx,\sigma)=-V(\sigma)+\alpha_c h(x)=-\dfrac{1}{2}\|\sigma(\bx)\|^2+\alpha_c h(x),
	\end{equation*}
	where $\alpha_c$ is  such that $h_c(\bx_0,\sigma(\bx_0))> 0$. We prove that $t\mapsto h_c(\bx(t),\sigma(\bx(t)))$ is increasing.  Notice first that the following holds:
	\begin{equation}
		\label{ec:ineqhcdot}
		\begin{split}
			\alpha_c \nabla h(x)\left(\sigma(\bx)+v(x)\right)&\geq -\alpha_c \eta \|\sigma(\bx)\|-\alpha_c \nabla h(x)v(x)\\
			&\geq -\alpha_c \eta \|\sigma(\bx)\|-\alpha_c \alpha h(x)\\
			&=-\alpha_c \eta \|\sigma(\bx)\|-\alpha \left(h_c(\bx) + V(\sigma)\right),
		\end{split}
	\end{equation}
	where the first line is deduced from $$\nabla h(x) \sigma(\bx) \geq  -\|\nabla h(x)\|\|\sigma(\bx)\|\geq -\eta \|\sigma(\bx)\|,$$ the second from $\nabla h(x)v(x)\geq -\alpha h(x)$ (by construction of $v$), and the third from the definition of $h_c$. The time-derivative of $h_c$ satisfies:	
	\begin{equation}
		\label{ec:hcdot}
		\begin{split}
			\dot h_c(\bx,\sigma)=&-\dot V(\sigma)+\alpha_c \nabla h(x)\dot x\\
			\geq& \kappa \|\sigma (\bx)\|+\alpha_c \nabla h(x )\left(\sigma(\bx)+v(x)\right)),\\
			\geq &\left(\kappa-\alpha_c \eta\right)\|\sigma (\bx)\|-{\alpha}V(\sigma)-\alpha h_c(\bx,\sigma )\\
			=&\frac{\alpha}{2}\|\sigma(\bx_0)\|\|\sigma(\bx)\|-\frac{\alpha}{2}\|\sigma(\bx)\|^2-\alpha h_c(\bx,\sigma)\\
			\geq &-\alpha h_c(\bx,\sigma),
		\end{split}
	\end{equation}
	where the second line follows from the fact that $\dot V(\sigma)\leq -\kappa \|\sigma(\bx)\|$, cf. proof of Theorem \ref{theo:safe_manifold}, and the definition of $\sigma(\bx)$; the third from  \eqref{ec:ineqhcdot}, the fourth from the definition of $\kappa$ in \eqref{ec:kappa} and the last one from \footnote{$\dot V(\sigma)\leq 0$ for any positive $\kappa$, then $\frac{1}{2}\|\sigma(x)\|^2=V(\sigma(x))\leq V(\sigma(x_0))=\frac{1}{2}\|\sigma(x_0)\|^2$} $\|\sigma(\bx_0)\|\geq \|\sigma(\bx)\|$.

	Since $h_c(\bx_0,\sigma(\bx_0))>0$, $\dot h_c(\bx,\sigma)\geq -\alpha h_c(\bx,\sigma)$  implies that $h_c(\bx(t),\sigma(\bx(t))\geq 0 $ for all $t\geq 0$. By definition of $h_c$, it holds that
	\begin{equation}
		\label{ec:ineq_th}
		h(x(t))\geq \frac{1}{\alpha_c}V(\sigma(\bx(t)))\geq 0 \:\forall t\geq 0. 
	\end{equation}
This finishes the proof.  \IEEEQEDclosed

The next corollary is directly deduced from from Theorem \ref{theo:safe_manifold} and  \eqref{ec:ineq_th}.
\begin{cor}
	Let  Assumption \ref{ass:delta} and Assumption \ref{ass:inputmatrix} be satisfied. System \eqref{ec:sys_di} in closed loop with  \eqref{ec:control_smc} and gain \eqref{ec:kappa} is safe with respect to $\mathcal{C}$. 
\end{cor}

A couple of comments are in order. First, notice that a safe reaching phase is ensured by the control \eqref{ec:control_smc} with \eqref{ec:kappa}.  We highlight that  achieving a safe reaching phase in the context of SMC is a non-trivial task. In \cite{Gomezancona2025}, it is proposed a planned trajectory-like solution to this problem. A closer look  makes evident that either if the initial condition is far from $\mathcal{S}$ or close to $\partial\mathcal{C}_1$ then $\kappa$ is large. As pointed out by the presented example later on, this control parameter  is conservative (large), since even with small values a safe reaching phase is achieved for initial conditions no far from the safe sliding domain.

Second, since there might be false equilibrium of the reduced order dynamics on the boundary of $\mathcal{C}_1$, it is not possible to guarantee global asymptotic stability of the closed-loop system. However, construction of safety and stabilizer control $v$ for simple integrator  can be adopted to construct the manifold $\mathcal{S}$; see, e.g. \cite{Reis2020control}.

	\subsection{Adaptive gain } 
\label{sec:barrierfunction}
While control \eqref{ec:control_smc} provides exact compensation of the perturbations on $\mathcal{S}$, it has  two drawbacks: (i) Chattering of the control signal, and (ii) requirement of the upper bound of the perturbation terms by Assumption \ref{ass:delta}. Next, we provide a barrier function-based gain to adjust the first and relax the second  \footnote{The concept of \textit{barrier function-based gain}, used in the adaptive framework \cite{Obeid2018barrier}, must not be confused with the concept of CBF used in the context of safety control.} .

For $\varepsilon>0$, let us define 
\begin{equation*}
	\mathcal{S}_\varepsilon:=\left\{\bx\in \mathcal{X}:\|\sigma(\bx)\|<\varepsilon\right\},
\end{equation*}
which corresponds to the so-called real sliding manifold, and the set
\begin{equation*}
	\mathcal{C}_\gamma:=\left\{ x\in \mathcal{X}_1: h_{\gamma}(x)=h(x)+\gamma\geq 0\right\},
\end{equation*}
where $\gamma>0$. Clearly, $\mathcal{C}_1\subset \mathcal{C}_{\gamma}$.

We consider control 
\begin{equation}
	\label{ec:control_bsmc}
	u(t,\bx)=k_{\varepsilon} (t,\bx)G^{-1}(\bx)\dfrac{\sigma(\bx)}{\|\sigma(\bx)\|},
\end{equation}
where
\begin{equation*}
	k_{\varepsilon}(t,\bx)=\left\{ \begin{array}{rr}
		k(t,\bx), & t\leq \tau \\
		k_{b}(\|\sigma(\bx(t))\|), & t>\tau.
	\end{array} \right.
\end{equation*}
Here $k(t,\bx)$ is given by \eqref{ec:gaink} with \eqref{ec:kappa}, $k_b(r)=\dfrac{r}{\varepsilon -r}$ and $\tau>0$ the smallest number such that $\|\sigma(\bx(\tau))\|\leq \frac{\varepsilon}{2}$. The gain $k_b$ was introduced as part of an adaptive SMC in \cite{Obeid2018barrier}.  The forthcoming result is build upon the next lemma, directly deduced from \cite{Obeid2018barrier}, \cite{cruz2023uniform}.
\begin{lem}
	\label{lem:barrier}
	Let Assumption \ref{ass:inputmatrix} be satisfied and system \eqref{ec:sys_di} be in closed loop with \eqref{ec:control_bsmc}. For any given $\varepsilon>0$, if $\|\sigma(\bx(\tau))\|\leq \frac{\varepsilon}{2}$ for some $\tau>0$, then $\|\sigma(\bx(t))\|<\varepsilon$ for all $t\geq \tau$.
\end{lem}

We next drop Assumption \ref{ass:delta}, at the expense of overestimating $\|\delta(t,x)\|$ with $d(t,\bx)$ in the gain $k(t,\bx)$.  We have the following result:
\begin{theo}
	Let Assumption \ref{ass:inputmatrix} be satisfied and for given $\gamma>0$ let $\varepsilon>0$ be such that $\varepsilon\leq \dfrac{\alpha \gamma}{\eta}$. 
	System \eqref{ec:sys_di} in closed loop with \eqref{ec:control_bsmc}  is safe with respect to $\mathcal{C}=\mathcal{C}_\gamma\times \mathcal{X}_2$. 
\end{theo}

\textit{Proof.} 
	According to the definition of the gain in \eqref{ec:control_bsmc}, there are two possible scenarios depending on the initial condition: (i) $\bx_0\notin \mathcal{S}_{\varepsilon/2}=\left\{\|\sigma(\bx_0)\|\leq \frac{\varepsilon}{2}\right\}$ or (ii) $\bx_0\in \mathcal{S}_{\varepsilon/2}$. Without loss of generality, let us assume the case (i), which implies that the control gain in \eqref{ec:control_bsmc} operates with $k(t,\bx)$ at the initial time. 
	
	%The proof is split into three parts: First, it is proved that, for any $\varepsilon>0$, the solution of \eqref{ec:sys_di}, \eqref{ec:control_bsmc} converges in finite time to the domain $\mathcal{S}_{\varepsilon/2}$; second, that before entering to $\mathcal{S}_{\varepsilon/2}$, it remains within $\mathcal{C}_\gamma$; and third, that once it achieves $\mathcal{S}_{\varepsilon}$ it remains therein for all $t\geq \tau$ and in turn on $\mathcal{C}_\gamma$.
	
	Notice first that any trajectory of \eqref{ec:sys_di} in closed loop with \eqref{ec:control_bsmc} converges  to the domain $\mathcal{S}_{\varepsilon/2}$. This follows from the right inequality in \eqref{ec:sigma_bound}. Indeed, 
	\begin{equation*}
		\|\sigma(\bx(t))\|\leq\frac{\varepsilon}{2},\: \text{for}\: t=\tau:= \dfrac{\sqrt{2}}{\kappa}\left(\|\sigma(\bx_0)\|-\frac{\varepsilon}{2}\right).
	\end{equation*}
	Hence, by Lemma \ref{lem:barrier}, $\|\sigma (\bx(t))\|\leq \varepsilon$ for all $t\geq \tau$, i.e. $\bx(t)\in \mathcal{S}_{\varepsilon}$ for all $t\geq \tau$. 
	We prove in two parts  that the system is  safe w.r.t. $\mathcal{C}$, equivalently that $x(t,\bx_0)\in \mathcal{C}_{\gamma}$ for all $t\geq 0$.
	
	1. For $t\in [0,\tau]$: Let us define
	\begin{equation*}
		h_{c\gamma}(\bx,\sigma)=-V(\sigma)+\alpha_c h_{\gamma}(x)=h_c(\bx,\sigma)+\alpha_c\gamma,
	\end{equation*}
	which by definition of $\alpha_c$ in Proposition \ref{propo:reaching} satisfies $h_{c\gamma}(\bx_0,\sigma(\bx_0))>0$, and adopt the same arguments from the proof of Proposition \ref{propo:reaching}. In fact, from \eqref{ec:hcdot},
	\begin{equation*}
		\begin{split}
			\dot h_{c\gamma}(\bx,\sigma)=&\dot h_c(\bx,\sigma)\\
			\geq& \left(\kappa-\alpha_c\eta\right)\|\sigma(\bx)\|-\dfrac{\alpha}{2}\|\sigma(\bx)\|^2-\alpha h_c(x,\sigma)\\
			=& \left(\kappa-\alpha_c\eta\right)\|\sigma(\bx)\|-\dfrac{\alpha}{2}\|\sigma(\bx)\|^2\\
			&-\alpha h_{c\gamma}(\bx,\sigma)+\alpha \alpha_c\gamma\\
			\geq& -\alpha h_{c\gamma}(\bx,\sigma),\:t\in [0,\tau],
		\end{split}
	\end{equation*}
	which implies that $h_{c\gamma}(\bx, \sigma)\geq 0$ whenever $x_0\in \mathcal{C}_\gamma $ and in turn that $x(t,\bx_0)\in \mathcal{C}_{\gamma}$ for $t\in [0,\tau]$.
	
	2. For $t\geq \tau$:  That  $\bx(t)\in \mathcal{S}_{\varepsilon}$ for all $t\geq \tau$ is equivalent to 
	\begin{equation*}
		\dot x=v(x)+\xi(\bx), \:\|\xi(\bx(t))\|<\varepsilon,\:\forall t\geq \tau.
	\end{equation*}	
	Hence, for all $t\geq \tau$,
	\begin{equation*}
		\begin{split}
			\dot h_{\gamma}(x)=&\nabla h(x)\dot x\\
			=&\nabla h(x)v(x)+\nabla h(x)\xi,  \\
			\geq& -\alpha h(x)-\eta\varepsilon\\
			=&-\alpha (h_\gamma(x)-\gamma)-\eta\varepsilon\\
			\geq &-\alpha h_{\gamma}(x)+\alpha \gamma -\eta\varepsilon.
		\end{split}
	\end{equation*}
	Since $\varepsilon\leq \dfrac{\alpha \gamma}{\eta}$, then  $\dot h_{\gamma}(x)\geq -\alpha h_{\gamma}(x)$, and $h_{\gamma}(x(t))\geq 0$, i.e. $x(t,\bx_0)\in \mathcal{C}_{\gamma}$, for all $t\geq \tau$. \IEEEQEDclosed

	\section{Numerical simulations}
\label{sec:example}

Let system \eqref{ec:sys_di} with $n=2$ and $G(\bx)=I$. We consider the CBF
\begin{equation*}
	h(x)=\|x-\xun\|^2-r^2,
\end{equation*}
where $\xun\in \mathbb{R}^2$ is the center position of an unsafe zone  characterized by a circle of radius $r$, so that the safe set is $$\mathcal{C}_1=\left\{x\in\mathcal{X}_1: h(x)=\|x-\xun\|^2-r^2\geq 0  \right\},$$ where $\mathcal{X}_1=[-3,6]\times [0,6]$. Let  $\vdes(x)=-(x-x_{goal})$, with  $x_{goal}=\begin{pmatrix}
	3 & 5
\end{pmatrix}^T$. Safety controller follows from \eqref{ec:usafe_gen} with $\nabla h(x)=2 (x-\xun)^T$ and $\alpha=1$. We smooth $\vsafe$ by the function $\vsafe^s$ with $s=0.5$ and simulate the disturbances with functions $\Delta_b(t,\bx)=0.25\sin(x) \bold{1}_{2\times 2}$ and $\delta(t,\bx)=5\sin(5t)\bold{1}_{2\times 1}$. The obstacle is considered to have center
$
\xun=\begin{pmatrix}
	2 &3
\end{pmatrix}^T$ and radius $r=1$.

\begin{figure}[htbp!]
	\centering
	\includegraphics[scale=0.8]{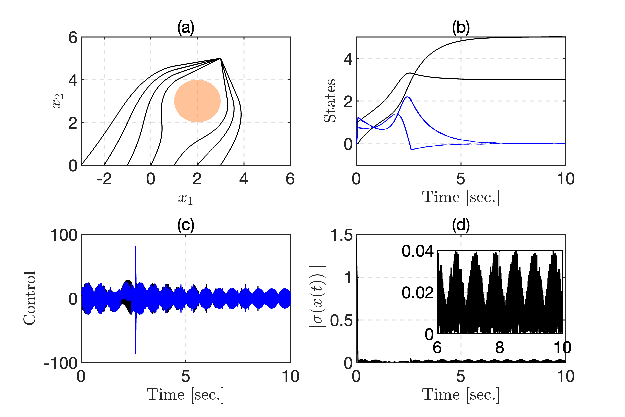}
	\caption{(a) Position trajectories of \eqref{ec:sys_di} in closed loop with \eqref{ec:control_smc} in the plane $(x_1,x_2)$, for several initial conditions. Light orange corresponds to $\mathcal{X}_1\setminus\mathcal{C}_1$; (b) States evolution in time for the initial condition $\begin{pmatrix}
			1 & 0
		\end{pmatrix}^T$: position are in black and velocities in blue; (c) Control signals for the initial condition $\begin{pmatrix}
			1 & 0
		\end{pmatrix}^T$; (d) Sliding variable evolution in time for the initial condition $\begin{pmatrix}
			1 & 0
		\end{pmatrix}^T$. }
	\label{fig:simulation}
\end{figure}

\paragraph{Safe SMC} The control \eqref{ec:control_smc} is constructed with gain \eqref{ec:gaink} and \eqref{ec:kappa}, where $\beta=0.1$, $\eta=6.4031$, $d(t,\bx)=\|\delta(t,\bx)\|$ and $\mu=-0.5$. Figure \ref{fig:simulation} displays  the simulation results. For $\bx_0=\begin{pmatrix}
	1 & 0
\end{pmatrix}^T$, $\kappa=4.0277$, but no reported simulations show that a safe reaching phase holds for values less than that. The overshoot of the control signal around $t=2.5$ seconds is induced by small values of the parameter $s$ of  the smooth function $\nu^s$.

\begin{figure}[htbp!]
	\centering
	\includegraphics[scale=0.8]{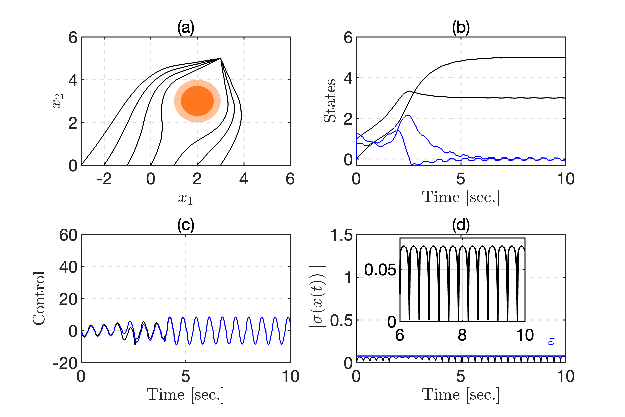}
	\caption{(a) Position trajectories of \eqref{ec:sys_di} in closed loop with \eqref{ec:control_smc} in the plane $(x_1,x_2)$, for several initial conditions. Light orange corresponds to $\mathcal{X}_1\setminus\mathcal{C}_1$ and deep orange to $\mathcal{X}_1\setminus\mathcal{C}_\gamma$; (b) States evolution in time for the initial condition $\begin{pmatrix}
			1 & 0
		\end{pmatrix}^T$: position are in black and velocities in blue; (c) Control signals for the initial condition $\begin{pmatrix}
			1 & 0
		\end{pmatrix}^T$; (d) Sliding variable evolution in time for the initial condition $\begin{pmatrix}
			1 & 0
		\end{pmatrix}^T$. }
	\label{fig:simulation2}
\end{figure}

\paragraph{Adaptive gain} We now simulate control \eqref{ec:control_bsmc} by setting $\gamma=0.5$, which gives $\varepsilon=0.0781$. We suppose the same input perturbation $\Delta_b(t,\bx)$ and
\begin{equation*}
	\delta(t,\bx)=\left\{\begin{array}{cc}
		5\sin(10t)\bold{1}_{2\times 1} & t\leq4\\
		9\sin(10t)\bold{1}_{2\times 1} & t>4.\\
	\end{array}  \right.
\end{equation*}
At the initial time, the gain $k_\varepsilon(t,\bx)$ is considered to be $k(t,\bx)$ with constant $d(t,\bx)=20$, which overestimate the bound on $\delta$, supposed to be unknown. The simulation results are depicted in Figure \ref{fig:simulation2}.

\section{Conclusion}
\label{sec:conclusion}
We introduced two robust control algorithms for critical-safety position of systems represented by a double integrator within the framework of sliding mode control. The proposed approach serves as the starting point for research in several directions, e.g. safe higher-order sliding mode control and safe sliding mode control with input saturation. Moreover, they might be particularly relevant, for instance, for robust obstacle avoidance in robotics when the model of the robot is not fully known \cite{Molnar2021model}. In the presence of multiple obstacles, one can generalize the approach by always selecting the nearest obstacle in the design of $v$ at any given moment, which gives rise to a nonsmooth CBF \cite{Glotfelter2020nonsmooth}.

	\bibliographystyle{unsrt}
	\bibliography{SSM_double_integrator}  

\end{document}